\documentclass[aps,prl,reprint,
superscriptaddress,
 amsmath,amssymb,
 aps,
]{revtex4-2}
\usepackage{graphicx,hyperref}
\usepackage{dcolumn}
\usepackage{bm}

\begin{document}

\preprint{APS/123-QED}

\title{SAFT-P: A Plaquette-Level Perturbation for Self-Assembly in Patchy Colloids}

\author{Hamza \c{C}oban}
\email{hamzaco@mit.edu}
\affiliation{%
 Department of Physics, Massachusetts Institute of Technology, Cambridge, Massachusetts 02139, United States
}%
\author{Alfredo Alexander-Katz}%
 \email{aalexand@mit.edu}
\affiliation{%
 Department of Materials Science and Engineering, Massachusetts Institute of Technology, Cambridge, Massachusetts 02139, United States
}%

\date{\today}

\begin{abstract}
We introduce SAFT-P, a plaquette-level extension of Statistical Associating Fluid Theory for patchy particles. By treating local clusters as associating superparticles and contracting their free energy back to monomer densities, SAFT-P retains information about patch topology that is lost in conventional SAFT. Grand-canonical Monte Carlo simulations of binary and ternary mixtures show that SAFT-P captures topology-dependent critical points and coexistence curves and discriminates between particles with identical valence but different patch layouts. Beyond topology, incorporating plaquette-scale correlations also improves predictions in regimes where patch specific interactions are absent. Results indicate that resolving correlations at the plaquette scale provides an analytical route to model complex condensates and self-assembly with topology-sensitive local structure.
\end{abstract}

\maketitle

Multivalent macromolecules and patchy particles with directional binding can undergo liquid–liquid phase separation and self-organize into (biomolecular) condensates or colloidal networks whose rheology and connectivity are governed by binding valence, specificity, and orientation constraints.\cite{banani2017biomolecular,brangwynne2017,alberti2019considerations,bianchi2011patchy,Zhang2004}. In biological settings, this idea is supported by cellular examples where multivalency drives phase transitions and enables switch-like control of signaling and organization \cite{li2012phase,su2016phase}. At the level of coarse-grained statistical mechanics, a central message from the patchy-particle literature is that valence alone is not sufficient: interaction specificity, steric constraints, conformation changes and patch placement bias, thereby shifting coexistence boundaries and the onset of aggregation or gelation \cite{Dorsaz2012,zhu2020self,Audus2018valence}. Recent imaging studies further reveal heterogeneous network architectures within biomolecular condensates, consistent with nonuniform local connectivity and packing that emerge from directional and multivalent interactions \cite{zhou2025}.

In effectively two-dimensional environments, such as membrane-bound or surface-tethered systems, reduced coordination number and strong excluded-volume effects make local bonding geometry and patch arrangement particularly important in determining connectivity and phase behavior.\cite{russo2010association,litschel2024membrane,hsu2023surface}. Indeed, membrane- or surface-associated condensates provide concrete realizations of effectively two-dimensional phase separation in signaling contexts, where stoichiometry and local connectivity regulate both composition and function \cite{lin2022two,case2019stoichiometry}. Taken together, these studies motivate theoretical approaches that can distinguish particles or macromolecules with identical valence but different bonding geometry, since such differences can control the preferred local motifs and, consequently, the global phase behavior \cite{zhu2020self,bianchi2011patchy}.

This sensitivity to microscopic bonding rules has motivated sustained efforts to develop tractable theories across scales, from mean-field mixtures to network-based and field-theoretic descriptions. To model interaction networks and reproduce the equilibrium phase behavior of biomolecular condensates, a range of frameworks have been developed across multiple coarse-graining levels and resolutions \cite{shimobayashi2021nucleation,saar2023theoretical,jacobs2023theory}. Mean-field descriptions of multicomponent liquids, such as Flory–Huggins–type models, provide a baseline and yield qualitative—and in some settings semi-quantitative—predictions for phase diagrams and multiphase coexistence \cite{Mao2019,van2023predicting,zwicker2022evolved}. Continuum phase-field (Cahn–Hilliard–type) approaches extend these free-energy descriptions to spatiotemporal dynamics, including binodal/spinodal evolution and coarsening, and can incorporate hydrodynamic effects to examine coarsening laws and pattern selection \cite{Shrinivas2021}. A sticker-and-spacer (or linker) framework, represents molecules as collections of binding sites connected by flexible segments and can capture sequence-encoded multivalency, percolation/gelation, and coupling between network formation and demixing; within this viewpoint, Flory–Stockmayer (FS) theory offers predictions for gelation thresholds and cluster-size statistics in step-growth/associative networks \cite{choi2020physical,harmon2017intrinsically}. Within the stickers-and-spacers paradigm, while some treatments emphasize effectively irreversible network growth, the Rubinstein–Semenov theory of associating polymers provides an analytical description of reversible network formation \cite{semenov1998thermoreversible}.
 
Coarse-grained “patchy-particle” models provide a minimalist route to encode multivalency and directional bonding. Developed on Wertheim's thermodynamic perturbation theory (TPT), Statistical Associating Fluid Theory (SAFT) provides an analytical framework that has seen tremendous success as a molecularly grounded equation-of-state for associating fluids across diverse applications, including electrolyte solutions, polymer fluids, surfactant systems, supercritical fluids, and hydrogen-bonding liquids \cite{muller2001}. Building on this foundation, SAFT has also been proven successful in various applications in patchy colloids \cite{bianchi2011patchy,teixeira2017phase,Audus2018valence} and biomolecular condensates \cite{sanders2020competing}. This framework is effective since it encodes valence and binding topology explicitly, allowing quantitative predictions of multi-phase coexistence, rigorous assessment of how multivalency and sticker patterning shift phase boundaries, and rational design of interaction motifs to engineer condensate behavior \cite{espinosa2020liquid,dai2023engineering}.

SAFT provides a molecular theory for associating fluids and has been widely used for quantitative predictions of equilibrium properties. Higher-order extensions (TPT2, TPT3) are available \cite{shukla2000tpt2,vlvcek2003thermodynamic,economou2002statistical}, and multibody/cluster-based corrections have been introduced to extend SAFT beyond the independent-site assumptions of TPT1—for example to treat arbitrary bonding-site geometries and situations where association involves strong multibody correlations \cite{zhu2020self}. Yet standard SAFT treatments remain insensitive to the spatial arrangement of bonding sites at fixed valence \cite{jacobs2014phase,Teixeira2019} even though, that arrangement can be altered by geometric stereochemistry, e.g. cis/trans isomers or enantiomers, which changes which bonds can form simultaneously and which local motifs are accessible, thereby shifting assembly pathways and phase behavior \cite{shikha2025inversion,pi2021efficient}. Ring corrections which depends on Monte Carlo integration have been proposed \cite{Sear1994,Marshall2013}, but they are not universal: they depend on the assumed bonding topology and site model, typically restricting the theory to a small subset of associating site pairs, fixed valence and geometry, and a limited class of cyclic graphs (frequently only small rings or cyclic dimers/double-bond motifs), rather than providing an unrestricted treatment of multicomponent plaquettes of arbitrary composition.

In this Letter, we introduce SAFT-P, an extension of SAFT that computes equilibrium properties at the cluster level to capture phase behavior in monomeric patchy-particle fluids. While the SAFT-P construction is general and can be formulated for other cluster choices and dimensions, here we use 2×2 plaquettes on a two-dimensional square lattice as the minimal cluster that resolves patch geometry and local packing constraints. Our approach provides a general treatment of 2D biomolecular condensates and protein mixtures, with sufficient resolution to capture spatial patch constraints. Across a set of binary and ternary mixtures, SAFT-P yields phase diagrams that are in closer agreement with simulation than conventional SAFT. Moreover, because SAFT-P resolves local patch geometry, it predicts phase separation between isomers, an effect that standard SAFT cannot capture.

Beyond mean‑field treatments, SAFT offers a perturbative correction by systematically incorporating short‑range correlations and directional bonding through a graph‑theoretic cluster expansion \cite{Wertheim1984,economou2002statistical,zmpitas2016detailed}. 
Explicitly accounting for directional bonding, the first-order residual Helmholtz free energy due to association is given by:
\begin{equation}
    \beta a_{\text{res}} = \left(\ln x - x + \frac{1}{2}\right), \quad x = \frac{\rho_0}{\rho},
\end{equation}
where $x$ is the fraction of unbounded (monomer) species. This formalism explicitly distinguishes monomer contributions from bonded species and provides a robust approach to model fluids with directional interactions.\\

In a generic many-component system, each particle has some fixed number of patches through which specific bonds occur. An isotropic interaction energy arises between any pair of neighbors regardless of the orientations of the patches. The free energy density per site of such a system can be written as the sum of the ideal entropic contribution and directional/anisotropic interaction terms. Here, we do not introduce empty sites since we add a non-interacting solvent as a different species in our framework and we assume the whole system is incompressible.
\begin{equation}
\begin{aligned}
f ={} \sum_i \phi_i \ln\phi
_i + \sum_i \phi_i \sum_s \left( \ln X_i^s - \frac{X_i^s}{2} + \frac{1}{2}\right),
\end{aligned}
\end{equation}
where $\phi$ is the total particle fraction in a lattice,  and $\phi_i$ the fraction of species $i$, $X_i^s$ is the unbounded fraction of type $s$ patches on type $i$ component. The first term is the entropic term and the latter is due to directional patch bondings noting that nondirectional interactions may be embedded into directional setting. The fractions $X_i^s$ forms a coupled equation system which generally does not have a closed-form solution
\begin{equation}
    X_i^s= \left(1+\sum_j \phi_j \sum_{s'} m_j^{s'} X_j^{s'} \Delta^{ss'}\right)^{-1},
    \label{eq:fractions}
\end{equation}
where the first sum is over all unique particle types and the second sum is over unique patches on type $j$ particles. The number of type $s$ patches on a type $j$ molecule is given with $m_j^{s'}$. Lastly, $\Delta^{ss'}$ is the association strength of patches $s$ and $s'$. For the lattice setting, it has a simpler expression, namely, $z\Delta^{ss'}=e^{\beta \epsilon_{ss'}}-1$, where $\epsilon_{ss'}$ is the bonding energy between the patches. The extension to the continuum is straightforward: the lattice-specific $\Delta^{ss'}$ is replaced by the corresponding association integral for off-lattice patchy spheres \cite{Wertheim1984, almarza2012three}. \\

\begin{figure}[h!]
    \centering
    \includegraphics[width=0.5\linewidth]{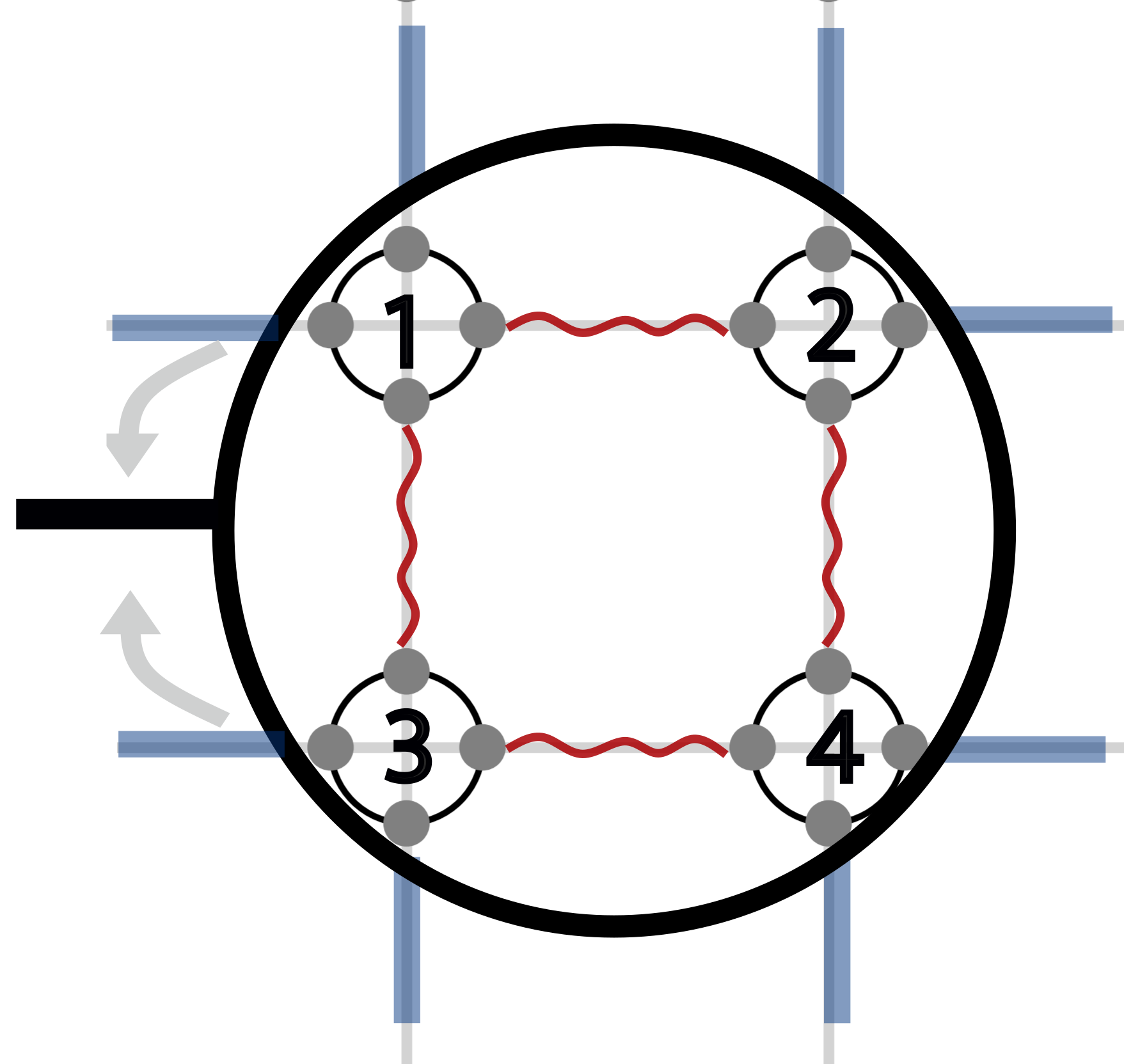}
    \caption{A schematic of a $2\times2$ plaquette and conversion of it into a superparticle. Internal bonds are in red and boundary edges in blue. Edge pairs in the same direction is converted to a single super-edge.}
    \label{fig:placeholder}
\end{figure}
To capture condensation substructures and distinguish patchy particles that share the same valence but differ in patch arrangement, we use a plaquette-level coarse graining. We model small clusters of monomers as superparticles by mapping each $(2\times2)$ lattice plaquette—containing four primitive monomers—onto an effective particle with four superpatches. Each superpatch comprises two colocalized patches on the same plaquette edge and exhibits cooperative bonding: the two patches are treated as binding (or unbinding) simultaneously. To avoid double counting while retaining geometric resolution, we partition plaquette microstates into equivalence classes defined by their outgoing edge pattern (i.e., the set of external patches presented to neighboring plaquettes). Within each class, we integrate out the internal degrees of freedom—constituent identities and internal bonds—by performing a Boltzmann-weighted average over all microstates in the class to define a single canonical (effective) plaquette. The resulting free energy per lattice site for a mixture of plaquettes is then
\begin{equation}
\begin{aligned}
    \bar{f} =& \frac{1}{4}\left(f_{\text{entropy}} + f_{\text{association}} +f_{\text{intra-bonds}} \right) \\
    =&{}\frac{1}{4}\sum_p \psi_p  \left(\ln\frac{\psi_p}{g_p} \ + \sum_s m_p^s \left( \ln X_p^s - \frac{X_p^s}{2} + \frac{1}{2}\right) + \mu_p\right) 
\end{aligned}
\end{equation}
where the factor of 4 accounts the number of sites in a plaquettes and ensures that $f$ remains the free energy per site, $\psi_p$ is the fraction of plaquette $p$ and $g_p$ accounts for rotational multiplicity/degeneracy of plaquette $p$. $X_p^s$ is the unbounded fraction of type $s$ super-patches on type $p$ plaquette. The non-directional and directional bonds within plaquettes and the chemical potential of constituent monomers are captured by the last term. Specifically,
\begin{equation}
\begin{aligned}
    &\mu_p = \sum_{i=1}^4 \mu_i +\varepsilon_{12}+\varepsilon_{24}+\varepsilon_{13}+\varepsilon_{34},\\ &\varepsilon_{ij}=\varepsilon_{ij,d} + \varepsilon_{ij,nd}
\end{aligned}
\end{equation}
where $\mu_i$ is the chemical potential of the monomer $i$ and $\varepsilon_{ij}$ is the total interaction coupling between monomer $i$ and $j$ including the directional (d) and non-directional (nd) terms.

Given $n$ monomer species, we enumerate the distinct $2\times 2$ plaquettes and obtain $N$ unique plaquette classes after accounting for degeneracies. Let $\vec{\psi}\in\Delta^{N}$ denote the plaquette-fraction vector and $\vec{\phi}\in\Delta^{n}$ the monomer-fraction vector. The linear map $C:\Delta^{N}\to\Delta^{n}$ relates these compositions by counting the monomer constituents of each plaquette, i.e., $\vec{\phi}=C\vec{\psi}$. The monomer-level free-energy density is then obtained by contracting the plaquette free energy,
\begin{equation}
    f(\vec{\phi})=\inf_{\vec{\psi}\in\Delta^{N}:\ C\vec{\psi}=\vec{\phi}} \,\bar f(\vec{\psi}),
\end{equation}
which is the standard contraction of a higher-dimensional variational problem onto the monomer composition space \cite{touchette2009large,varadhan2010large,ellis2012entropy}.
In practice, this requires minimizing $\bar f(\vec{\psi})$ over $N$ plaquette fractions subject to the simplex constraints and the linear composition constraint $C\vec{\psi}=\vec{\phi}$. We perform this constrained optimization with a first-order method (AdaGrad), which is stable and scales to large $N$ in our setting. The main practical limitation is scaling with chemical complexity: for $n$ monomer species, the number of distinct $2\times 2$ plaquettes grows as $O(n^4)$ (before symmetry reduction), increasing the contraction dimension accordingly.

Moving on to case studies, since topology matters already in pure solutions of single species, we first quantify its impact by generating patchy–solvent phase diagrams of two-patch particle with different patch organizations. We then turn to a few-component mixture where patch geometry induces an isomer-isomer separation.

\begin{figure}[h!]
\centering
\makebox[\columnwidth]{%
  \includegraphics[width=0.49\columnwidth]{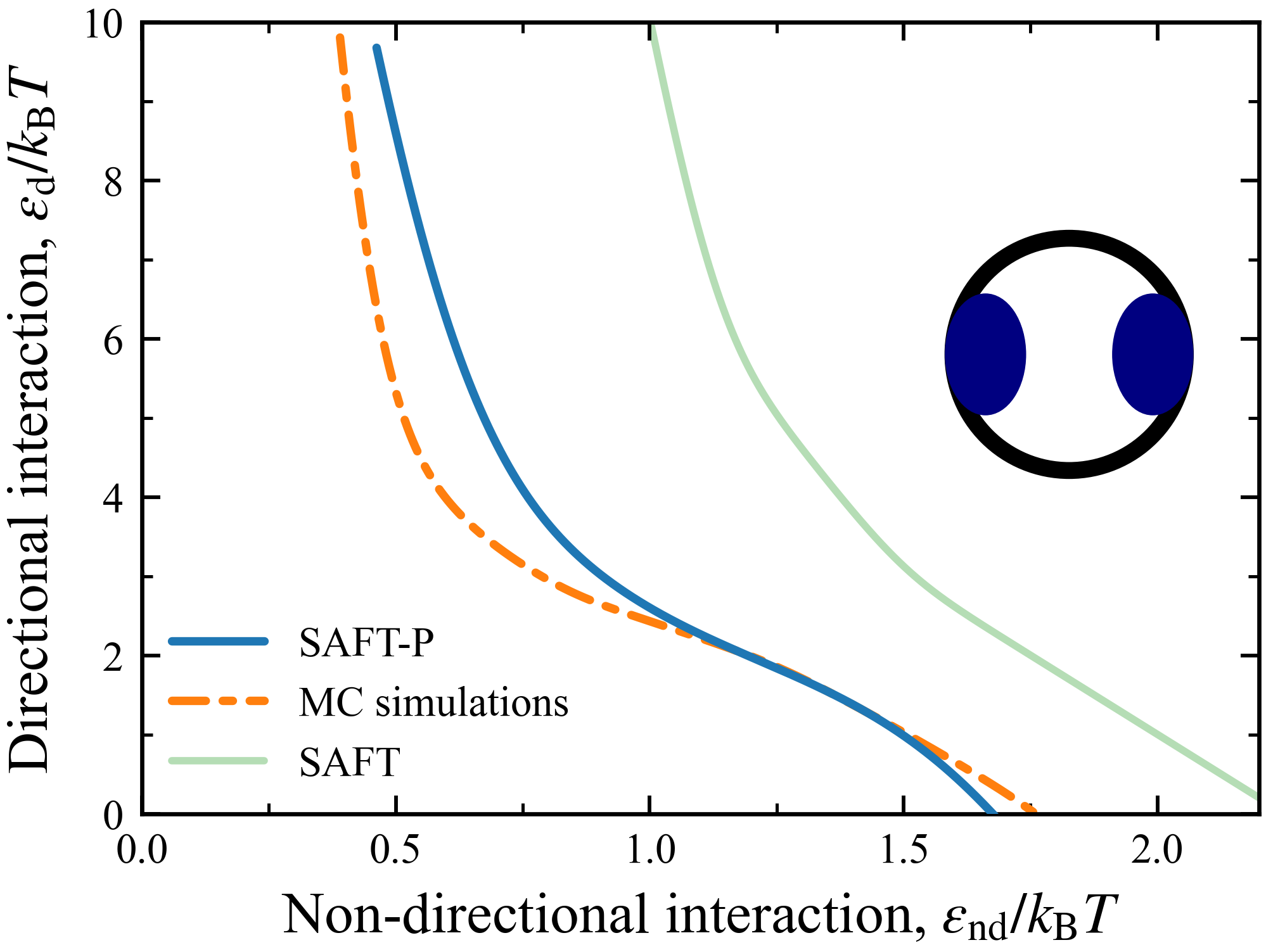}\hfill
  \includegraphics[width=0.49\columnwidth]{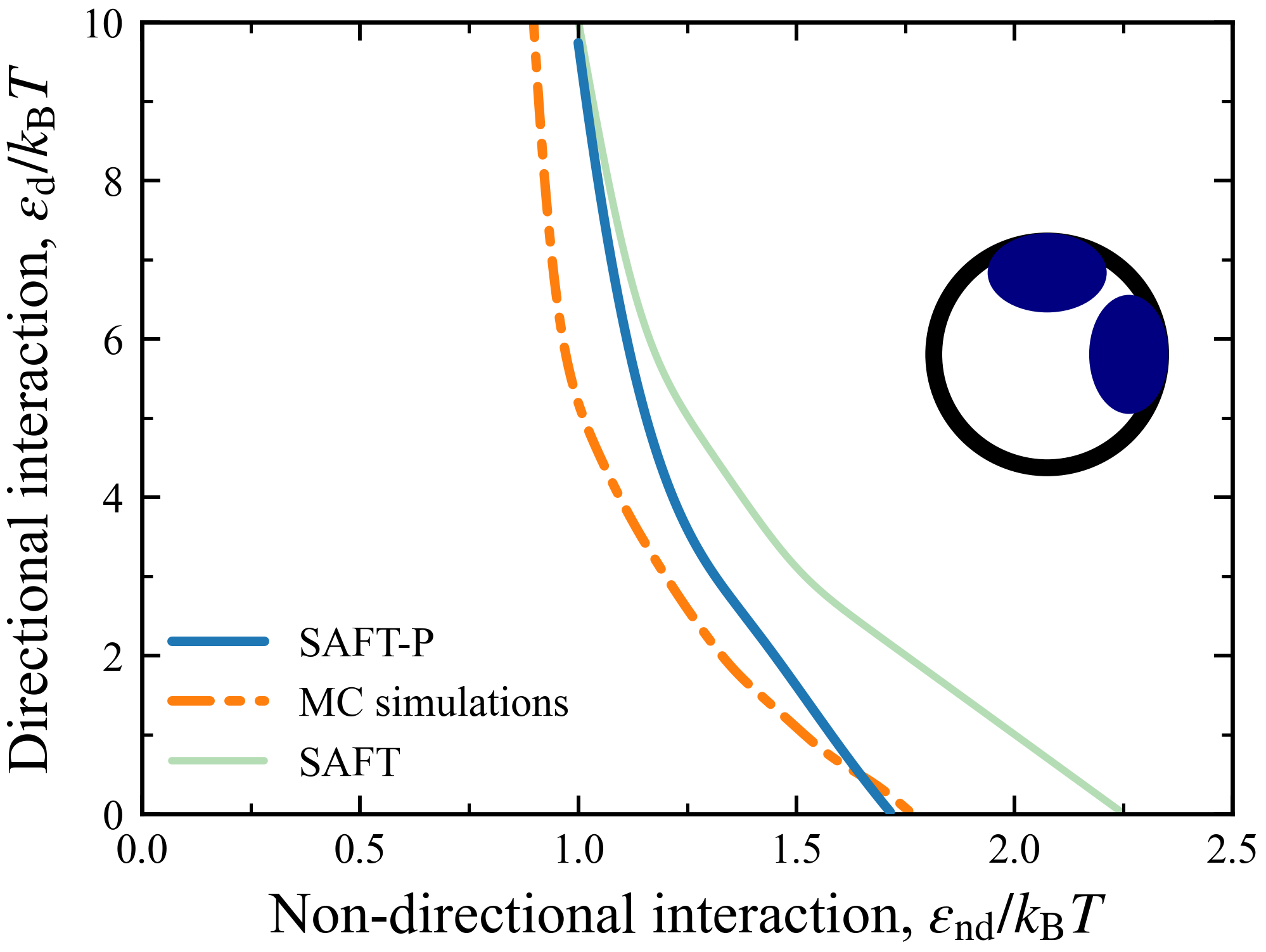}%
}
\caption{Critical lines for stick-shaped (left) and L-shaped (right) particles as a function of directional and non-directional interaction strengths; insets depict the corresponding patch geometries.}
\label{fig:critical_lines}
\end{figure}
Single-component patchy-particle solutions with identical patches are the simplest systems that still exhibit geometric (spatial) constraints. We study two-patch particles on a square lattice with either a $90^\circ$ patch arrangement (L-shaped) or a $180^\circ$ arrangement (stick-shaped). Patch–patch interactions are directional, while particle–particle interactions are non-directional and apply only between non-solvent particles. We performed grand-canonical ($\mu VT$) Monte Carlo simulations on a $10\times10$ square lattice using Metropolis sampling, with random identity change of patchy particles by defining each rotational degrees of freedom and empty sites as individual species without introducing any sampling bias. From simulations, we trace the critical line using mixed-field order parameters and tuning the chemical potentials for coexistence with histogram-reweighting techniques \cite{ferrenberg1988,jacobs2014phase}, exploiting the mapping of binary lattice mixtures into the Ising universality class \cite{wilding1993critical}. Within SAFT/SAFT-P, we locate the spinodal by the loss of convexity of the free energy $f(\phi)$, i.e., when the smallest eigenvalue of the Hessian with respect to species densities vanishes. We then compute the binodal via a spinodal-seeded (spinodal-assisted) common-tangent construction, using the spinodal points as initial bracketing/initial guesses for the coexistence search.

Monte Carlo simulations provide a numerically exact route to equilibrium properties and are sufficient, in principle, to determine coexistence and criticality. In practice, however, producing an accurate phase diagram is substantially more costly and methodologically involved than a single equilibrium run: one must ensure equilibration across many state points, tune fields/chemical potentials to enforce coexistence, verify histogram overlap for reweighting, control finite-size effects, and account for critical slowing down near the critical region. The resulting workflow is implementation-heavy and can be opaque to reproduce, as outcomes depend on the specific choice of order parameters, biasing/reweighting strategy, and tuning protocol. By contrast, SAFT-P trades sampling for a deterministic free-energy construction: phase boundaries follow from convexity loss and common-tangent conditions, enabling systematic parameter sweeps without additional coexistence tuning or statistical noise.

In Fig.~\ref{fig:critical_lines}, we compile critical points across interaction settings and delineate the resulting critical line in the $(\varepsilon_{\mathrm{nd}},\varepsilon_{\mathrm{d}})$ plane. Below this boundary the system remains single phase, whereas above it phase separation occurs for appropriate compositions and chemical potentials. Standard SAFT captures the critical line for L-shaped particles reasonably well but shows larger deviations for stick-shaped particles. We attribute this difference to particle geometry: L-shaped particles have greater rotational freedom and can form more isotropic, globular bonding networks, mitigating the impact of SAFT’s simplifying assumptions. In contrast, the stick geometry constrains bonding and promotes more anisotropic network growth, for which standard SAFT is less accurate. SAFT-P improves on standard SAFT by explicitly resolving the dominant local motif for sticks—stacked configurations within $(2\times2)$ plaquettes. As a result, SAFT-P not only reproduces the critical line for stick-shaped particles but also shows that the leading instability is driven by fluctuations in the population of stacked plaquettes, i.e., incipient aggregation of these stacked motifs. 

\begin{figure}[h!]
    \centering
    \includegraphics[width=0.5\linewidth]{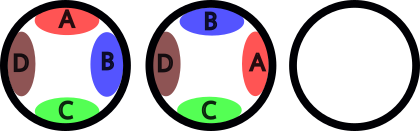}
    \caption{Components of the proposed isomeric mixture, From right to left: geometric isomer 1, geometric isomer 2, solvent}
    \label{fig:components}
\end{figure}

Geometric isomers can exhibit distinct phase behavior in designed or spatially constrained environments. Capturing this sensitivity requires a treatment beyond standard SAFT, because the free energy depends not only on composition but also on patch geometry and directional bonding patterns. Here we introduce a minimal ternary lattice model consisting of two geometric isomers and an inert solvent. The patch layouts of the two isomers and the solvent are shown in Fig.~\ref{fig:components}. We include directional attractions between complementary patch pairs A–C and B–D, together with like–like self-attractions C–C and D–D, choosing $(\epsilon_{CC}=\epsilon_{DD}\equiv\epsilon_s)$ and $(\epsilon_{AC}=\epsilon_{BD}=1.5 \epsilon_s)$. This hierarchy favors the designed A–C/B–D bonds; consequently, an isolated isomer embedded in a region rich in the other isomer cannot simultaneously satisfy its preferred bonding pattern, creating local bonding frustration at the interface and within mixed domains.

\begin{figure}[h!]
    \centering
    \includegraphics[width=0.7\linewidth]{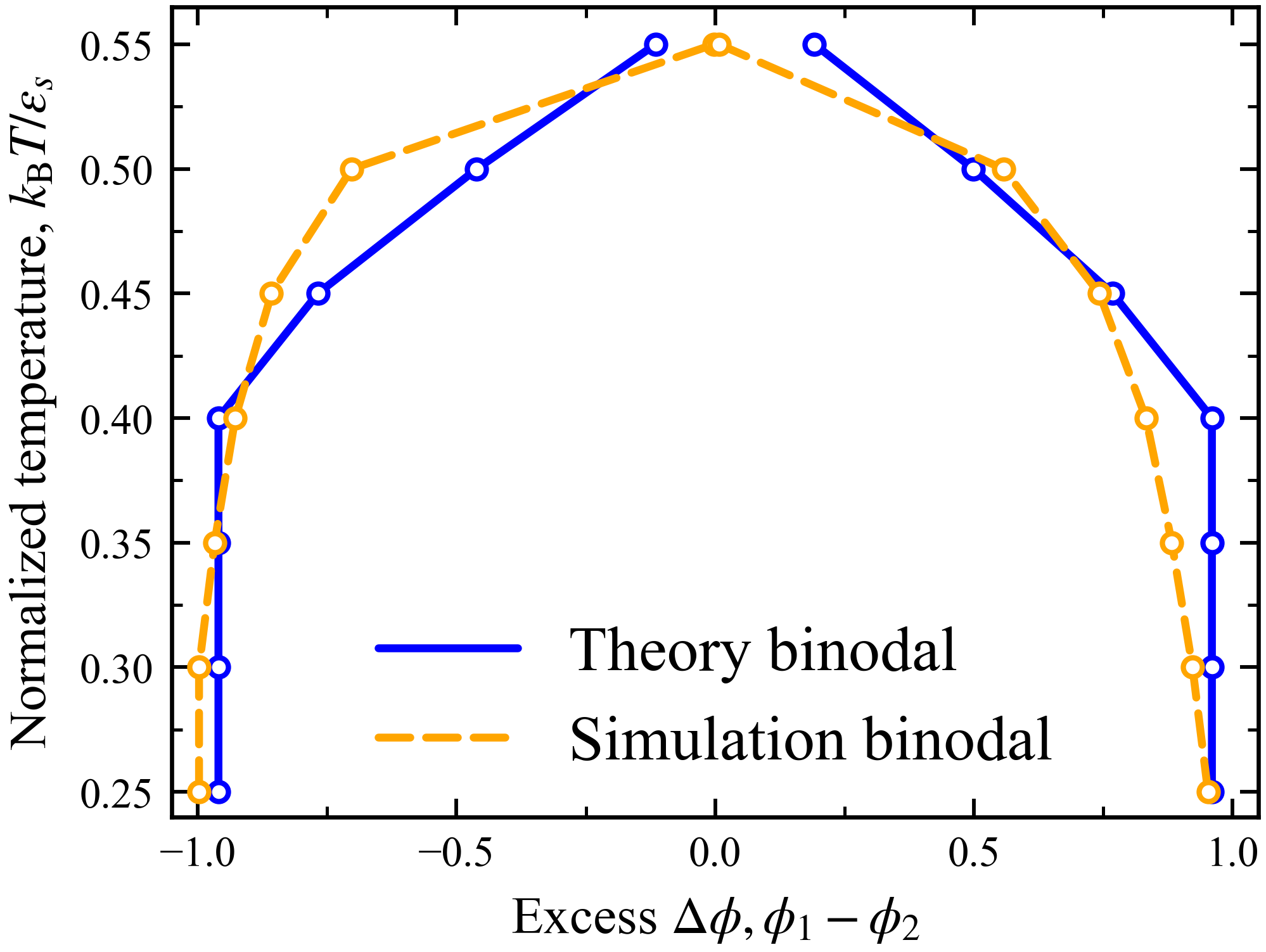}
    \caption{Binodal of the isomeric mixture. Coexistence curve plotted as normalized temperature versus excess composition, where $\phi_1,\phi_2$ are the isomeric fractions in the two coexisting phases. Solid blue: theoretical binodal from plaquette coarse-grained free-energy / SAFT-P. Dashed orange: simulation binodal from Monte Carlo simulations on an $20\times20$ lattice }
    \label{fig:placeholder}
\end{figure}

Because the solvent is noninteracting, changing its fraction mainly dilutes the mixture and does not qualitatively change the coexistence physics, aside from finite-size effects. We therefore fix the solvent fraction to $1/2$, which reduces the problem to a one-dimensional composition scan. On the theory side, we compute the SAFT-P free energy along this cut and obtain the binodal via a common-tangent construction. Monte Carlo simulations are performed in a mixed canonical–grand-canonical setup: the solvent is treated canonically (its number is fixed), while the two isomeric components are sampled grand canonically at fixed (zero) chemical potentials. In the two-phase regime the distribution of the isomeric excess becomes bimodal; we identify the peak locations and use them to construct the binodal from simulation.

Figure~\ref{fig:placeholder} compares the coexistence curve obtained from SAFT-P with Monte Carlo results. The two binodals show close agreement across the scanned compositions, and SAFT-P reproduces the critical temperature of the ternary mixture on this finite lattice. In contrast, standard SAFT does not distinguish the two geometric isomers at fixed valence and interaction strengths and therefore cannot capture the excess-composition binodal. These results emphasize that incorporating plaquette-scale correlations resolves the geometry-induced bias in local bonding and its thermodynamic consequences in a minimal isomeric mixture.

We introduced SAFT-P, a plaquette-level extension of SAFT that retains local correlation and patch-topology information by treating $2\times2$ clusters as associating superparticles and contracting their free energy back to monomer densities. Across single- and few-component lattice mixtures, SAFT-P systematically improves phase-boundary predictions relative to standard SAFT: it captures the dominant stacked motif governing the instability of stick-shaped particles and correctly resolves geometric-isomer separation in a minimal ternary mixture where conventional SAFT remains topology-blind. More broadly, these results show that incorporating a small amount of local structure into an analytic associating-fluid framework can reconcile tractability with topology sensitivity, providing a route to model designed patchy mixtures and membrane-like two-dimensional condensates. Extensions to larger clusters, additional interaction motifs, and higher-component mixtures follow naturally within the same construction.

\bibliography{biblio}
\end{document}


\maketitle
\section{Methods}

\subsection{Constructing plaquettes}

In the main text, we did not spell out the bookkeeping needed to treat each $2\times 2$ plaquette as an effective single species (“superparticle”). A plaquette contains four monomers, each with up to four patches. Bonds formed among these monomers generate up to four distinct intra-plaquette couplings, depending on which neighboring faces are occupied by compatible patches. We incorporate these internal bonds by assigning an effective chemical potential to the plaquette species. Concretely, if a plaquette configuration has total internal bonding energy $\epsilon$, we set its plaquette chemical potential to include this contribution (i.e., $\mu_{p}\rightarrow \mu_{p}+\epsilon)$. In this way, $\mu_{p}$ represents the free-energy cost of inserting that plaquette into the system, and it naturally decomposes into the sum of constituent monomer chemical potentials plus the internal bonding energy of the plaquette.

\begin{figure}[h!]
\centering
\includegraphics[width=0.2\linewidth]{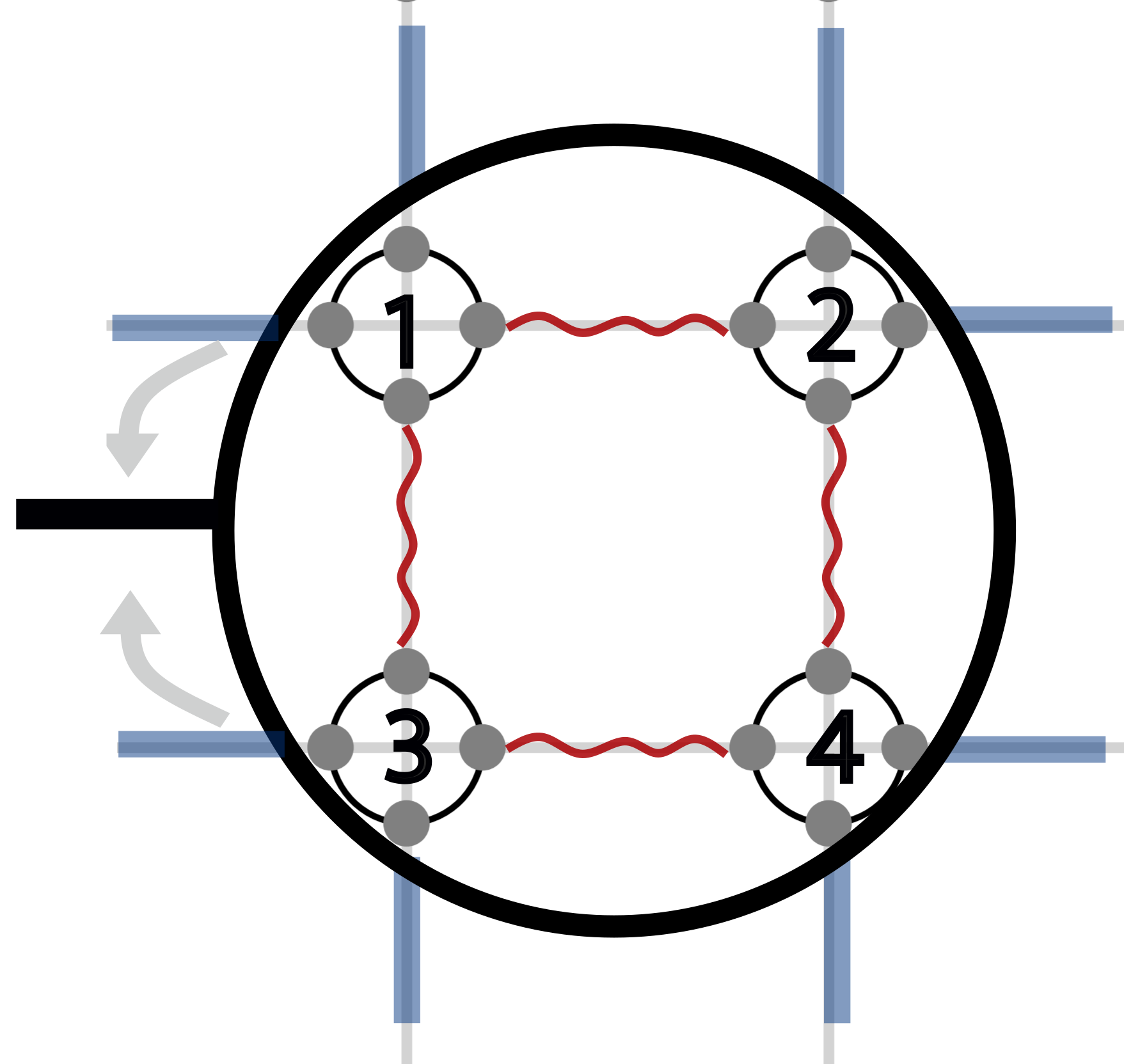}
\caption{A $2\times 2$ plaquette treated as a single patchy particle (superparticle).}
\label{fig:placeholder}
\end{figure}

To handle the “outgoing” (inter-plaquette) patches, we introduce superpatches: on each side of a plaquette, the two boundary monomer patches are grouped into a single effective patch. The key approximation is cooperative bonding on a side: the two constituent boundary patches are not treated as independently satisfiable; instead, the side is either bonded as a unit or unbonded as a unit. With this convention, the interaction between two facing superpatches is defined as the sum of the underlying patch–patch interactions across the interface. For example, if plaquette 1 presents patches (A) and (B) on one side, and the opposing side of plaquette 2 presents complementary patches (C) and (D), then the superpatch–superpatch interaction energy is

$$\epsilon_{\mathrm{super}} = \epsilon_{AC} + \epsilon_{BD}$$

with the ordering fixed by the microscopic arrangement of patches along the side. This ordering matters because distinct plaquette microstates can contain the same set of monomers yet differ in the spatial ordering of their outgoing patches, leading to different inter-plaquette bonding energies.

\subsection{Plaquette microstates and notation}
A plaquette microstate is a $2\times 2$ configuration of corner species
\begin{equation}
\mathbf{s}=(s_{\mathrm{UL}},s_{\mathrm{UR}},s_{\mathrm{BL}},s_{\mathrm{BR}}),
\end{equation}
with $(\mathrm{UL},\mathrm{UR},\mathrm{BL},\mathrm{BR})$ a fixed corner order. Each species $s$ carries four directional patch labels
\begin{equation}
\texttt{patches}[s]=(p_N,p_E,p_S,p_W),
\end{equation}
and patch--patch interaction energies are specified by a matrix $J$.

The intra-plaquette energy is the sum of the four bonds internal to the $2\times 2$ block,
\begin{equation}
\begin{aligned}
E(\mathbf{s})=&\;
J\!\big(p^{\mathrm{UL}}_{E},p^{\mathrm{UR}}_{W}\big)
+J\!\big(p^{\mathrm{BL}}_{E},p^{\mathrm{BR}}_{W}\big)\\
&+J\!\big(p^{\mathrm{UL}}_{S},p^{\mathrm{BL}}_{N}\big)
+J\!\big(p^{\mathrm{UR}}_{S},p^{\mathrm{BR}}_{N}\big).
\end{aligned}
\end{equation}
When plaquette classes merge across different compositions, we include chemical potentials through
\begin{equation}
\mu(\mathbf{s})=\mu_{s_{\mathrm{UL}}}+\mu_{s_{\mathrm{UR}}}+\mu_{s_{\mathrm{BL}}}+\mu_{s_{\mathrm{BR}}},
\qquad
E_{\mathrm{eff}}(\mathbf{s})=E(\mathbf{s})-\mu(\mathbf{s}).
\end{equation}

\subsection{Boundary-edge representation (the class label)}
To couple plaquettes to neighbors we retain only the outgoing boundary information. With four perimeter ``edges'' (top, right, bottom, left), each edge is encoded as an edge ID that combines the two patch labels exposed on that side. Concretely,
\begin{equation}
\mathbf{b}(\mathbf{s})=
\big(b_{\mathrm{top}},b_{\mathrm{right}},b_{\mathrm{bottom}},b_{\mathrm{left}}\big),
\end{equation}
with
\begin{equation}
\begin{aligned}
b_{\mathrm{top}}    &= \mathrm{id}\!\big(p^{\mathrm{UL}}_{N},p^{\mathrm{UR}}_{N}\big),\\
b_{\mathrm{right}}  &= \mathrm{id}\!\big(p^{\mathrm{UR}}_{E},p^{\mathrm{BR}}_{E}\big),\\
b_{\mathrm{bottom}} &= \mathrm{id}\!\big(p^{\mathrm{BR}}_{S},p^{\mathrm{BL}}_{S}\big),\\
b_{\mathrm{left}}   &= \mathrm{id}\!\big(p^{\mathrm{BL}}_{W},p^{\mathrm{UL}}_{W}\big),
\end{aligned}
\end{equation}
and $\mathrm{id}(\cdot,\cdot)$ is a one-to-one encoding of a patch pair (optionally symmetrized if edges are treated as undirected).

We define plaquette equivalence classes by matching boundary-edge patterns up to rotation: two plaquettes belong to the same class if their boundary-edge tuples $\mathbf{b}$ are identical after a global rotation by $0^\circ,90^\circ,180^\circ,$ or $270^\circ$. Operationally, each class is labeled by the lexicographically minimal rotated version of $(b_{\mathrm{top}},b_{\mathrm{right}},b_{\mathrm{bottom}},b_{\mathrm{left}})$.

\subsection{Boltzmann-weighted coarse-graining within a class}
Let $g$ denote a boundary-edge class. The class contains many microstates $\mathbf{s}$ that share the same outgoing boundary edges but differ in internal composition and internal bonding. We integrate out these internal degrees of freedom by a Boltzmann sum over the class:
\begin{equation}
Z_g = \sum_{\mathbf{s}\in g}\exp\!\big[-E_{\mathrm{eff}}(\mathbf{s})\big].
\end{equation}
This defines an effective statistical weight for the class and implies an effective free energy
\begin{equation}
F_g = -\ln Z_g,
\end{equation}
(up to the usual choice of units/absorbed $\beta$).

Any internal observable $A(\mathbf{s})$ is represented at the class level by its conditional Boltzmann average
\begin{equation}
\langle A\rangle_g = \frac{1}{Z_g}\sum_{\mathbf{s}\in g} A(\mathbf{s})\,e^{-E_{\mathrm{eff}}(\mathbf{s})}.
\end{equation}

In practice we store two coarse-grained quantities for each class:
\begin{enumerate}
\item \textbf{Average species composition} (used to build the constraint rows $A\rho=\text{targets}$):
\begin{equation}
\langle n_\alpha\rangle_g
=\frac{1}{Z_g}\sum_{\mathbf{s}\in g} n_\alpha(\mathbf{s})\,e^{-E_{\mathrm{eff}}(\mathbf{s})},
\end{equation}
where $n_\alpha(\mathbf{s})\in\{0,1,2,3,4\}$ counts how many corners are occupied by species $\alpha$.

\item \textbf{Average boundary-edge multiplicities} (used in the association term):
\begin{equation}
\langle m_e\rangle_g
=\frac{1}{Z_g}\sum_{\mathbf{s}\in g} m_e(\mathbf{s})\,e^{-E_{\mathrm{eff}}(\mathbf{s})},
\end{equation}
where $m_e(\mathbf{s})\in\{0,1,2,3,4\}$ counts how many of the four perimeter edges in $\mathbf{b}(\mathbf{s})$ equal edge type $e$.
\end{enumerate}

This construction produces one \emph{canonical plaquette} per boundary-edge class $g$: it has a fixed outgoing edge pattern (the class label), while its internal composition and internal bonding are represented thermodynamically through Boltzmann-weighted averages.

\subsection{Summary of SAFT-P Calculations}

In the Letter, we introduced a free-energy formulation in plaquette space and used a contraction principle to obtain the corresponding free energy in monomer space. In practice, we discretize monomer space on a grid; each grid point defines a distinct constraint that selects the admissible plaquette compositions consistent with the prescribed monomer fractions. For each constraint, we define the optimization objective as the plaquette-space free energy evaluated on the constrained manifold and minimize it by gradient descent. This yields the contracted free energy at that monomer-space grid point.

Because the contracted free energy is estimated numerically, it exhibits small sampling/optimization noise that can induce spurious structure in finite-difference derivatives. We therefore apply LOESS (locally estimated scatterplot smoothing) to the contracted free-energy surface before differentiating. The resulting smooth free energy enables stable evaluation of the Hessian (second-derivative) matrix, whose eigenstructure identifies thermodynamic instabilities. We use the vanishing of the appropriate Hessian eigenvalue to locate critical points and assemble the resulting critical line.

\subsection{Monte Carlo simulations}

We simulate a two-dimensional patchy-particle lattice model on an $L\times L$ square lattice with periodic boundary conditions. Each lattice site is occupied by a species label (including an optional vacancy/solvent species). Each species carries a four-component patch vector corresponding to its faces $(N,E,S,W)$, and nearest-neighbor interactions are defined by a patch--patch energy matrix $J$. For any nearest-neighbor bond, the energetic contribution is the matrix element $J_{\alpha\beta}$ for the two facing patches $\alpha$ and $\beta$. Periodic boundary conditions are enforced by modulo indexing in both lattice directions. 

Dynamics are generated by Metropolis Monte Carlo updates. At each attempted update, the code proposes a local change of the lattice configuration and accepts it with probability
\begin{equation}
p_{\mathrm{acc}}=\min\left(1, e^{-\beta \Delta \mathcal{H}}\right),
\end{equation}
where $\beta$ is the inverse temperature and $\Delta\mathcal{H}$ is the change in the (effective) Hamiltonian between the proposed and current configurations. The energy difference is computed locally from the bonds incident to the updated site(s), using the patch identities on the relevant faces and the interaction matrix $J$. 

To support grand-canonical control of selected species, the simulation uses a per-species chemical potential table. A designated index set (``$\mu$-coupled species'') is assigned chemical potential $\mu$, while all other species have $\mu=0$. The simulation records the corresponding $\mu$-controlled population,
\begin{equation}
N_{\mu}=\sum_{i=1}^{N_{\text{sites}}} \mathbf{1}\{\sigma_i \in \mathcal{S}_{\mu}\},
\qquad
\rho_{\mu}=N_{\mu}/N_{\text{sites}},
\end{equation}
where $\sigma_i$ is the species at site $i$ and $\mathcal{S}_{\mu}$ is the $\mu$-coupled set. These quantities are returned as time series at the sampling cadence described below. 

At fixed sampling intervals, the simulation records (i) species densities and derived total density, (ii) an energy-like observable $u$ (the per-site energetic contribution consistent with the bond-count convention used in the run), and (iii) bond-count observables that partition nearest-neighbor contacts into ``specific'' and ``non-specific'' contributions. Specifically, each undirected nearest-neighbor bond is counted once by scanning only the right and down bonds on the periodic lattice. This yields the total occupied--occupied contact count $C_{\mathrm{tot}}$. A bond is classified as ``specific'' if both facing patches equal a designated patch ID (implemented as patch ID $1$), producing $C_{\mathrm{sp}}$. The remaining contacts are labeled non-specific, $C_{\mathrm{ns}}=C_{\mathrm{tot}}-C_{\mathrm{sp}}$.

\subsection{Summary of Reweighting Techniques}

This workflow implements a sequential “parameter-tracking” procedure that maps a coexistence line of a two-species lattice model (here, two patchy particle species) by repeatedly adjusting the thermodynamic control parameters so that the order-parameter fluctuations match a universal critical reference distribution. At each step, one interaction parameter is advanced (the specific interaction strength $\epsilon_{\mathrm{sp}}$, and the remaining parameters (the non-specific interaction strength $\epsilon_{\mathrm{ns}}$ and the chemical potential $\mu)$ are re-tuned so the system remains at the same fluctuation fixed point. The result is a set of triplets $(\epsilon_{\mathrm{ns}}^*, \epsilon_{\mathrm{sp}}, \mu^*)$ that trace a path through parameter space consistent with the chosen universal target.

A universal reference distribution is constructed from the two-dimensional Ising model at criticality. The code runs a Wolff cluster algorithm at the exact critical temperature $(T_c=2/\ln(1+\sqrt{2}))$ (in units $J=k_B=1)$ and collects magnetization samples after burn-in. These samples are standardized to a zero-mean, unit-variance variable $x$, and a smoothed histogram $P^*(x)$ is produced by Gaussian convolution. This $P^*(x)$ serves as the target probability density function against which the lattice mixture’s suitably defined order parameter is matched. Using the Ising fixed-point distribution is standard in finite-size scaling analyses because it fixes the nonuniversal mixing between energy-like and density-like operators while preserving universal shape information.

Because direct simulation at the exact coexistence chemical potential is expensive, the code uses histogram reweighting in $\mu$ to locate $\mu^*$ near equal-phase weights. Given samples collected at a reference $\mu_0$, it computes importance weights $w(\mu)=\exp[-\beta(\mu-\mu_0)N_\mu]$ (up to normalization) and searches for a $\mu^*$ such that the reweighted density distribution has equal probability mass in predefined “dilute” and “dense” density windows. This is implemented as a bracketing/bisection routine driven by the signed difference of the integrated histogram density in the two windows, while monitoring the effective sample size (ESS) as a reliability diagnostic. Once $\mu^*$ is identified, the code either directly uses the importance weights or draws a systematic resample from the weighted ensemble to obtain approximately unweighted samples at $\mu^*$.

To connect the mixture’s observables to the Ising reference distribution, the code performs mixed-field finite-size scaling. It defines an order-parameter-like variable $M=\rho_\mu - su$, where $s$ is a mixing coefficient that accounts for the nonuniversal rotation between density-like and energy-like directions in the space of fluctuating fields. For a given $s$, the code centers and normalizes $M$ (using weighted mean and variance) to form a standardized variable $x$, then estimates its distribution by a fixed-bandwidth KDE on the same $x$-grid as the Ising reference. It optimizes $s$ (and a nuisance parameter $r$ that is penalized but otherwise not used downstream) by minimizing a combined mismatch objective: an $L^2$ difference to $P^*(x)$ plus an antisymmetry penalty about $x=0$ to enforce the expected symmetry at the fixed point. This step determines the appropriate field mixing so that the mixture’s fluctuations are compared to the correct universal scaling operator.

Finally, at each step in $\epsilon_{\mathrm{sp}}$, the code adjusts $\epsilon_{\mathrm{ns}}$ and $\mu$ to keep the standardized order-parameter distribution aligned with $P^*(x)$. It constructs a reweighting objective that compares the per-bin probability mass of the model’s $x$-histogram to the Ising reference mass on identical bin edges, using the Jensen–Shannon divergence as a stable, symmetric discrepancy measure. The trial reweighting weights are computed from the sampled bond counts and particle numbers via an exponential factor $\exp[-\Delta(\beta\epsilon_{\mathrm{ns}})C_{\mathrm{ns}} - \Delta(\beta\epsilon_{\mathrm{sp}})C_{\mathrm{sp}} - \Delta(\beta\mu)N_\mu]$ (with $\epsilon_{\mathrm{sp}}$ fixed to the new target value in that step), and the optimizer searches locally around the previous solution to find updated $\epsilon_{\mathrm{ns}}^*$ and $\mu^*$ that minimize the divergence. The resulting updated parameters are then used as the reference for the next increment in $\epsilon_{\mathrm{sp}}$, producing a continuous tracked path through parameter space.

In summary, the code combines (i) a universal Ising fixed-point distribution obtained by cluster Monte Carlo, (ii) grand-canonical lattice simulations of the mixture returning extensive observables suitable for histogram reweighting, (iii) $\mu$-reweighting to enforce equal-phase weights, (iv) mixed-field scaling to identify the correct order parameter, and (v) local reweighting-based optimization to retune $\epsilon_{\mathrm{ns}}$ and $\mu$ as $\epsilon_{\mathrm{sp}}$ is swept. The practical outcome is an efficient continuation method for tracing a coexistence/critical manifold without re-simulating from scratch at every parameter point, while maintaining direct contact with universal finite-size scaling constraints.